\documentclass[conference]{IEEEtran}
\IEEEoverridecommandlockouts
\usepackage{xurl}
\usepackage[hidelinks]{hyperref}

\usepackage{cite}
\usepackage{makecell}
\usepackage{amsmath,amssymb,amsfonts}
\usepackage{algorithmic}
\usepackage{graphicx}
\usepackage{textcomp}
\usepackage{xcolor}
\usepackage{tabularx}
 
\def\BibTeX{{\rm B\kern-.05em{\sc i\kern-.025em b}\kern-.08em
    T\kern-.1667em\lower.7ex\hbox{E}\kern-.125emX}}
\begin{document}

\title{MeAJOR Corpus: A Multi-Source Dataset for Phishing Email Detection}

\author{\IEEEauthorblockN{Paulo Mendes}
\IEEEauthorblockA{\textit{GECAD, ISEP} \\
\textit{Polytechnic of Porto}\\
4249-015 Porto, Portugal \\
\url{https://orcid.org/0009-0006-6060-772X}}
\and
\IEEEauthorblockN{Eva Maia}
\IEEEauthorblockA{\textit{GECAD, ISEP} \\
\textit{Polytechnic of Porto}\\
4249-015 Porto, Portugal \\
\url{https://orcid.org/0000-0002-8075-531X}}
\and
\IEEEauthorblockN{Isabel Praça}
\IEEEauthorblockA{\textit{GECAD, ISEP} \\
\textit{Polytechnic of Porto}\\
4249-015 Porto, Portugal \\
\url{https://orcid.org/0000-0002-2519-9859}}
}

\maketitle

\begin{abstract}
Phishing emails continue to pose a significant threat to cybersecurity by exploiting human vulnerabilities through deceptive content and malicious payloads. While Machine Learning (ML) models are effective at detecting phishing threats, their performance largely relies on the quality and diversity of the training data. This paper presents MeAJOR (Merged email Assets from Joint Open-source Repositories) Corpus, a novel, multi-source phishing email dataset designed to overcome critical limitations in existing resources. It integrates 135894 samples representing a broad number of phishing tactics and legitimate emails, with a wide spectrum of engineered features. We evaluated the dataset’s utility for phishing detection research through systematic experiments with four classification models (RF, XGB, MLP, and CNN) across multiple feature configurations. Results highlight the dataset's effectiveness, achieving 98.34\% F1 with XGB. By integrating broad features from multiple categories, our dataset provides a reusable and consistent resource, while addressing common challenges like class imbalance, generalisability and reproducibility.
\end{abstract}

\begin{IEEEkeywords}
phishing, dataset, email, artificial intelligence, machine learning, deep learning
\end{IEEEkeywords}

\section{Introduction}

Phishing remains one of the most prevalent and damaging forms of cybercrime, counting with over 3.76 million cyberattacks in 2024, causing wide financial losses globally and contributing to hundreds of thousands of compromised accounts (mostly email) in 2024~\cite{apwg_q1, apwg_q2, apwg_q3, apwg_q4}. It is also responsible for 15\% of attack vectors in data breaches, costing an average of \$4.88 million per breach~\cite{IBM_data_breach_report_2024}. Email is one of the many media through which these attacks are performed, as it remains a central channel for personal and professional communication, and with the ever-evolving attack landscape, automated detection and mitigation strategies have become essential to safeguarding users and infrastructures. Machine Learning (ML) approaches promise scalable and adaptive defence mechanisms against evolving phishing campaigns, however, their effectiveness is highly dependent on the quality and diversity of the training data.

Despite the availability of several open-source phishing datasets, recent literature highlights some limitations in their usability for robust model development~\cite{Mbadiwe2024}. These issues include extreme class imbalance, limited feature engineering, narrow coverage of phishing tactics, outdated samples, and inconsistent preprocessing practices~\cite{Mbadiwe2024, Alhuzali2025, AlSubaiey2024}. Furthermore, many studies rely on private or small-scale datasets, which limits generalisability and reproducibility~\cite{afonso2025rethinking}. These challenges restrain the development of efficient and realistic ML models, as they may lead to overfitting or inflated performance metrics on small or biased test sets.

In this paper, we address these limitations by analysing and comparing several existing open-source email phishing datasets to assess their strengths and limitations, as well as several related works (Section \ref{sec:related_works}). Based on this analysis, we propose combining multiple datasets to increase the diversity of phishing examples and expand the overall sample size, thereby enhancing the robustness and generalisability of ML models. We apply consistent preprocessing and feature engineering across various feature categories, resulting in a comprehensive and ready-to-use dataset tailored for ML applications, named MeAJOR (Merged email Assets from
Joint Open-source Repositories) Corpus (Section \ref{sec:dataset_compilation}).

To verify the utility of this dataset, we conduct different experiments using four distinct ML and Deep Learning (DL) models, each trained on different feature subsets (Section \ref{sec:experiments}). As is discussed in Section \ref{sec:discussion}, the results consistently demonstrate the effectiveness of the proposed dataset in supporting accurate and reliable phishing detection. In conclusion, by offering a standardised, feature-rich, and scalable resource, this dataset serves as the main contribution of our study and a valuable foundation for future research and development in phishing detection, as further detailed in Section \ref{sec:conclusion}.

\section{Related Work}
\label{sec:related_works}

\subsection{Data Sources}
Datasets are sets of data relevant to a specific topic. Data can be artificially generated or collected from different sources. Datasets are labelled if they contain a classifying feature. On the datasets that will be described, labels identify each email sample as benign or phishing but the labels can have other meanings depending on the dataset.

There are several publicly available email spam and phishing datasets, some of which have been used for ML-based email phishing detection~\cite{Salloum2022}. Spam datasets are commonly used in phishing detection studies~\cite{Khalid2024, Altwaijry2024} because phishing is a type of spam and these emails share many characteristics, such as being unsolicited, unwanted, deceptive, and attempting to solicit an action~\cite{Juneja2014}. Table \ref{tab:datasets_summary} provides an overview of these datasets, including their content type, label availability, and sample sizes.

\begin{table}[htbp]
    \caption{Open-Source Datasets for Phishing Detection}
    \scriptsize
    \begin{center}
    \begin{tabular}{|c|c|c|c|}
    \hline
    \textbf{Dataset} & \textbf{Content} & \textbf{Label} & \textbf{Sample Size} \\
    \hline
    Enron Corpus & Email & No & 517,401 \\ 
    \hline
    \makecell{Nazario Phishing\\Corpus} & Email phishing & Yes & 11,527 \\ 
    \hline
    SpamAssassin & Email spam & Yes & 6,047 \\
    \hline
    Nigerian Fraud & Email phishing & Yes & 3,975 \\ 
    \hline
    TREC-05 & Email spam & Yes & 92,189 \\ 
    \hline
    TREC-06 & Email spam & Yes & 37,822 \\ 
    \hline
    TREC-07 & Email spam & Yes & 75,419 \\ 
    \hline
    CEAS-08 & Email spam & Yes & 39,154 \\ 
    \hline
    Ling-Spam & Email spam & Yes & 2,893 \\ 
    \hline
    Email4S & Email phishing & Yes & 18,650 \\
    \hline
    \end{tabular}
    \end{center}
    \label{tab:datasets_summary}
\end{table}

The corpora available for phishing detection vary significantly in origin and content. They range from large, unlabelled corporate archives like the Enron Corpus~\cite{enron_source}, often used as a source of benign emails, to specialised collections containing only malicious samples, such as the Nazario Phishing Corpus~\cite{nazario_source} and Nigerian Fraud letters~\cite{nigerian_source}. Many were curated specifically for academic research and benchmarking, including the seminal Ling-Spam~\cite{ling_source}, the SpamAssassin corpus~\cite{spamassassin_source}, and collections from standardised competitive events like the TREC datasets~\cite{trec5source, trec6source, trec7source} and the CEAS-08~\cite{ceas8source}. More recent efforts focus on developing labelled, phishing-specific datasets, for example by extracting and curating samples from larger collections, as demonstrated by initiatives like Email4S~\cite{email4s_source}. 

Existing phishing email datasets face significant challenges that restrict the development of robust ML-based detection systems. Recent reviews highlight that many corpora suffer from data imbalance, outdated structures, and limited feature engineering. There is often a trade-off between scope and diversity; classic spam corpora like the SpamAssassin dataset offer high source diversity, whereas dedicated phishing collections, such as the Nazario Phishing Corpus, tend to be smaller and narrower in scope, lacking coverage of varied phishing tactics. These differences can heavily impact study results~\cite{Alhuzali2025}.

Furthermore, much of the current research relies on small public or inaccessible private datasets, which affects generalisability. Ethical concerns, such as the disclosure of sensitive or personally identifiable information, also make more difficult to acquire new datasets. Although some studies attempt to mitigate these issues by merging multiple corpora, these efforts often remain limited in scale, combining only two or three sources~\cite{Kyaw2024}. Addressing these quality issues is crucial for improving the reliability and generalisability of phishing detection research~\cite{AlSubaiey2024, Mbadiwe2024}.

\subsection{Phishing Email Features}
In the phishing email domain, researchers commonly extract features from six principal sources: email body text, embedded URLs, attachments, message headers, HTML structure, and external domain reputation, to expose both technical anomalies (e.g. mismatched “From” headers or IP-based links) and social-engineering cues (e.g. urgency words or deceptive phrasing)~\cite{Lee2025, Gascon2018, Gualberto2020, Valecha2022}.

\textbf{Email Body Text Features} contain free-form writing that can reveal many social-engineering tactics and tricks used by attackers to exploit the human factor of a system~\cite{Desolda2021, Gallo2024}. These cues are evident not only in the direct content of the email but also in its lexicality, syntax, and semantics. Some approaches used to extract these features from text include vectorisation methods such as Bag of Words or TF-IDF~\cite{Tawil2024}, counts of trigger words like “verify,” “password,” and “urgent”~\cite{Yasin2016}, readability scores such as Flesch–Kincaid and Gunning Fog~\cite{Mieronkoski2020}, and sentiment analysis to detect emotional manipulation~\cite{Xiang2025}. Together, these linguistic characteristics help identify deceptive emails crafted to manipulate recipients.

\textbf{Embedded URL Features} are crucial in identifying phishing emails, as attackers often use deceptive links to lure victims. These features typically include various metrics related to the URL’s lexical structure, such as the total length of the URL, the number of subdomains it contains, the presence of IP-literal hostnames (e.g. https://192.168.1.1/home), the directory depth (measured by counting the segments separated by “/”), token counts derived by splitting the URL on characters like “.”, “-”, and “\_”, digit‑to‑letter ratios and the overall character entropy. Beyond these lexical characteristics, other suspicious indicators include the use of dubious top-level domains, the presence of UTF-encoded characters, and the employment of excessive URL-shortening services~\cite{Joshi2019, Ahmed2022, Xie2022}. These combined URL features help in detecting potentially harmful links embedded in phishing emails.

\textbf{Attachment Features} play a vital role in detecting malicious content within emails, as attachments can carry malicious payloads or impersonate legitimate forms. These features include the number and types of attached files (e.g. executables, Office documents containing macros, and compressed archives), file‐size anomalies, discrepancies between the file extension and the MIME type, embedded macros or scripts, and the count of embedded images. Additionally, statistical flags for unusually large or zero-byte attachments and signature-based checks on macro patterns are effective tools for identifying concealed malware or credential-harvesting forms~\cite{Chen2023, Rudd2018}. Together, these features help uncover malicious attachments that aim to compromise the recipient’s security.

\textbf{Message Header Features} carry email's transport and routing metadata, which attackers often spoof or manipulate to conceal their identity or mislead recipients. Notable examples include the number and sequence of the \texttt{Received} header fields, which records each server the email passed through and creates a chain with timestamps (abnormal counts or irregular sequences in these hops may indicate suspicious routing or email relaying through unexpected servers); discrepancies between the \texttt{From} header field, which shows the sender claimed email; the \texttt{Return-Path} field, which specifies the destination for bounce messages; and the envelope sender (the actual sender address used during SMTP transmission), which can reveal inconsistencies or attempts to mask the sender identity. Authentication results from protocols like SPF (Sender Policy Framework), DKIM (DomainKeys Identified Mail), and DMARC (Domain-based Message Authentication, Reporting \& Conformance) also provide strong spoofing indicators by verifying if emails are actually from who they claim to be from. Additional suspicious indicators include anomalous domains in the \texttt{Reply-To} field (which specifies the address that replies will be directed to) or \texttt{List-Unsubscribe} field (used for newsletter opt-outs), as attackers might use these fields to redirect victims or avoid detection; and unusual timestamp discrepancies, such as impossible or inconsistent sending times across headers. By quantifying these signals, detection models can effectively learn to identify forged or illicit senders and flag potentially malicious emails~\cite{Kulkarni2020}.

\textbf{HTML Structure Features} are essential for analysing emails that include HTML content, as attackers often embed hidden forms, scripts, or obfuscated elements to deceive recipients. These features involve counting and examining HTML tags commonly exploited in phishing attempts, such as \verb|<form>| (which can collect user credentials), \verb|<iframe>| (often used to load external content invisibly), \verb|<script>| (which may execute malicious code), \verb|<img>| (which can track user interaction or load deceptive visuals), and inline CSS (frequently used to hide elements from the recipient). Key structural indicators include the depth of tag nesting (which can suggest attempts to hide content deep within complex layouts), the ratio of hidden elements (e.g. those using \texttt{display:none}) to visible ones, mismatches between anchor text and the actual \texttt{href} targets (which can disguise malicious links), and the presence of JavaScript event handlers like \texttt{onload} or \texttt{onclick} (which can trigger actions when the user interacts with the email). By extracting and analysing these elements, it becomes possible to uncover the client-side techniques used by phishing kits to bypass basic text-based detection methods~\cite{Yoon2024, Yoon2025}.

\textbf{Reputation Features} extend beyond the immediate content of an email to evaluate the trustworthiness of the attacker’s domains. Important reputation indicators include the domain’s registration age, time remaining until domain expiration, domain name registrar and domain nameserver metadata (which can reveal connections to known malicious actors or unreliable service providers), DNS record time-to-live values (which indicate the refresh rate of domain information), and the geographic location of the hosting IP address (adds further context, as certain regions may be associated with phishing or cybercrime). Additionally, it is crucial to check whether the host IP or Autonomous System Number (ASN) is on known Blacklists. Domains with short lifespans (newly registered or rapidly changing), use obscure top-level domains, or rely on hosting providers with lax abuse policies tend to have a strong association with phishing campaigns. These reputation features provide valuable signals for identifying potentially malicious domains used in attacks~\cite{Hao2016, Chiba2025}.

\subsection{Phishing Detection Techniques}
There are many techniques used for phishing email detection. Some of the most traditional ones include reputation-based methods, which involve assessing the trustworthiness of a sender or website based on historical data~\cite{Kang2007}; heuristic-based methods, which rely on flexible, experience-informed criteria~\cite{Prakash2010}; and rule-based methods, which classify content according to predefined rules~\cite{Altwaijry2024, Doshi2023}. Despite having consistent detection rates for known threats, these methods lack zero-day attack detection and are hard to maintain, which led researchers to search for novel approaches, often based on Natural Language Processing (NLP), ML, and DL.

Omotehinwa et al.~\cite{Omotehinwa2023} evaluated ensemble ML models for spam email detection, emphasising hyperparameter optimisation. They trained RF and Extreme Gradient Boosting (XGB) models on textual features extracted from the Enron-Spam corpus. XGB performed best with an F1 of 98.16\%, demonstrating the effectiveness of ensemble learning in identifying email threats.

Naswir et al.~\cite{Naswir2022} investigated how effectively lexical URL features can detect phishing emails. They extracted features like IP addresses in URLs, special characters (e.g. '@', '-', and ‘\%’), URL entropy, and length, alongside basic email metadata, from the Nazario Phishing Corpus and the legitimate Enron Corpus. By optimising a Support Vector Machine (SVM) with the Cuckoo Search algorithm, their method achieved 91\% accuracy, demonstrating the strong predictive power of URL-based features for phishing detection.

Khalid et al.~\cite{Khalid2024} introduced LogiTriBlend, a stacked ensemble model for phishing email detection, combining SVM, Logistic Regression, RF, and XGB. They used a small, imbalanced "Phishing Email Dataset" (4591 emails) and addressed this imbalance with SMOTE data augmentation. It comprises Enron Corpus, SpamAssassin, Nazario Phishing Corpus and other sources. The models were trained on textual features, vectorised using TF-IDF, Word2Vec, and Doc2Vec. The Doc2Vec-based features with LogiTriBlend achieved the best F1 (99.41\%). However, the study's generalisability might be limited due to the small size of the original dataset and its reliance solely on text features.
Chanis et al.~\cite{Chanis2024} also used ensemble methods for phishing detection, uniquely combining stylometric features (such as formatting and writing style) with content-based features. They trained separate classifiers for each feature type, then integrated them using a stacked ensemble. This approach captured both the email's meaning and writing style, achieving a 98.43\% F1 on a private dataset, outperforming content-only models by around 2.2\%.

Atawneh and Aljehani~\cite{Atawneh2023} compared various Deep Learning (DL) architectures (CNNs, RNNs, LSTMs) for phishing detection, using textual features from email bodies from the Nazario and SpamAssassin corpora. Their proposed BERT+LSTM hybrid model achieved a 99.55\% F1, outperforming other standalone models.
Kaushik et al.~\cite{Kaushik2023} proposed another hybrid DL model, combining LSTM and CNN layers, for phishing detection. Their model processes URL-based and textual features (transformed into 2D image-like matrices) from a balanced dataset, constructed by combining phishing URL data from PhishTank~\cite{PhishTank} and legitimate URL data from Alexa~\cite{AlexaTopSites} with additional email datasets from various public repositories. It achieved an F1 of around 99\%, demonstrating the efficacy of combining sequential and spatial feature learning.

Despite most previously mentioned studies using text-based features, other information can also be extracted and used as input for the models. For example, Zhang et al.~\cite{Zhang2016} developed a phishing detection framework that uses diverse features beyond just text. Their approach extracts text from embedded images via Optical Character Recognition (OCR), combines it with URL-based, web-based, and rule-based features. Using these varied inputs, they trained a two-stage Extreme Learning Machine (ELM) model on a dataset of over 22000 webpages collected from PhishTank, APWG, and legitimate sources such as Alexa’s top sites, achieving 99.04\% accuracy.

Zhang et al.~\cite{Zhang2025} used 79 static email features (e.g., sender IP, links, attachments) and textual embeddings to train a RF classifier for phishing detection. Their model achieved 99.97\% accuracy on a large, custom dataset (660985 emails), combining benign emails from Enron Corpus (517401 samples), Trec-07 (25220 samples), and SpamAssassin dataset (6952 samples); and malicious phishing samples from Nazario Phishing Corpus (9510 samples), Wooyun XSS dataset~\cite{WooyunXSSDataset} (168 samples), Trec-07 (50199 samples), and additional 49136 samples generated by summarising and simulating novel types of malicious email attacks that exploit email protocol vulnerabilities. While this dataset is extensive, its heavy reliance on the Enron Corpus for benign emails might introduce bias.

Altwaijry et al.~\cite{Altwaijry2024} compared DL models for phishing email detection, using only email body and subject content from Nazario Phishing Corpus and SpamAssassin. They proposed a lightweight 1D-CNN architecture (1D-CNNPD), enhanced with recurrent layers like Bi-GRU. Their best model (1D-CNNPD + Bi-GRU) achieved a 99.66\% F1, outperforming traditional ML models and comparable DL approaches such as THEMIS~\cite{Fang2019} and DeepAnti-PhishNet~\cite{vinayakumar2018deepanti} and demonstrating the effectiveness of CNNs combined with recurrent mechanisms for phishing detection from text.
Another recent approach by Koide et al.~\cite{Koide2024} introduced ChatSpamDetector, which uses Large Language Models (LLMs), specifically GPT-4, for phishing detection. They prompt GPT-4 directly with raw email text, allowing it to infer phishing characteristics contextually, rather than relying on engineered features. This zero-shot/few-shot approach achieved 99.70\% accuracy on a private balanced dataset, showing LLMs' ability to capture subtle phishing signals.

All these studies consistently show RF and XGB as top performers in phishing detection, highlighting their reliability and effectiveness for real-world applications involving structured email features.
\section{Dataset Compilation}
\label{sec:dataset_compilation}

From the previously described datasets, we selected the Nazario Phishing Corpus, Nigerian Fraud, TREC-05, TREC-06, and TREC-07 to be combined into a new dataset, aiming to enhance the diversity of phishing examples and increase the overall sample size. These datasets were chosen because they provide both the original raw email messages, which allow greater flexibility in feature extraction, and the corresponding classification labels, making them suitable for phishing detection tasks. Together, these sources resulted in a raw dataset comprising 220932 email samples.

\subsection{Feature Engineering}
Feature engineering is the process of extracting and transforming raw email data into a set of informative features that ML models can understand and learn from. These features can be numerical (e.g. the number of links in an email) or categorical (e.g. the type of attachments an email has). Based on our analysis of commonly extracted features in recent research and their respective mapping to key components of phishing emails (such as body content, embedded links, attachments, headers, HTML structure, and domain reputation), we selected a representative subset of features from each category. Table~\ref{tab:dataset_features} provides a summary of this selection.

\begin{table}[htbp]
    \centering
    \scriptsize
    \caption{Resulting Dataset Features}
    \label{tab:dataset_features}
    \begin{tabularx}{0.5\textwidth}{|c|X|}
        \hline
        \textbf{Feature Name} & \textbf{Feature Description} \\
        \hline
        source & Original dataset from where the content came from \\ 
        \hline
        sender & Hashed email address of the sender \\
        \hline
        sender\_domain & Email domain of the sender \\
        \hline
        receiver & Hashed email address of the receiver \\
        \hline
        receiver\_domain & Email domain of the receiver \\
        \hline
        date & Date and Time of when the email was sent (RFC 2822) \\
        \hline
        subject & The subject of the email \\
        \hline
        content\_types & The content type of the body \\
        \hline
        body & The email body \\
        \hline
        urls & URLs extracted from the email body \\
        \hline
        url\_count & Number of URLs present on the email \\
        \hline
        url\_length\_max & Maximum URL length \\
        \hline
        url\_length\_avg & Average URL length\\
        \hline
        url\_subdom\_max & Maximum subdomain count \\
        \hline
        url\_subdom\_avg & Average subdomain count \\
        \hline
        attachment\_count & Number of attachments present on the email \\
        \hline
        has\_attachments & 0 if it does not have any attachments or 1 if it does \\
        \hline
        attachment\_types & The types of the attachments present on the email \\
        \hline
        language & The language (e.g. "en") of the email text \\
        \hline
        label & 0 for benign or 1 for phishing \\
        \hline
    \end{tabularx}
\end{table}

\subsection{Data Cleaning}
With the features properly defined and the unified dataset comprising 220932 email samples, the subsequent and crucial step to ensure the quality and reliability of our analysis is data cleaning. This process is fundamental, especially when combining distinct phishing datasets from various sources, as the heterogeneous nature of emails can introduce noise, which would negatively impact the development of robust models. Email messages come in diverse formats, character encodings, and content types. So, to preserve the original content, our pipeline is designed to transform each raw email message into a consistent form while maintaining privacy safeguards.

We begin by interpreting the header fields (sender, receiver, date, and subject) by decoding various character encodings into legible text. This ensures that international characters (e.g. Chinese, Arabic, Greek, and Cyrillic) and special symbols are accurately rendered, preserving the nuances of the original message metadata.

Emails often consist of multiple parts, each with its own content type and body, as already noted. These parts may be encoded using formats such as \texttt{Base64} or \texttt{quoted-printable}, requiring individual handling. During preprocessing, any content transfer encoding is automatically detected and decoded. HTML markup is then stripped to isolate the underlying textual content, resulting in a clean and continuous stream of words suitable for further analysis.

Many email threads include quoted replies and forwarded content. Our pipeline locates common reply separators and removes duplicate segments, isolating unique elements. By focusing on individual content, we reduce noise and highlight the elements most relevant to phishing behaviour. Beyond text, emails often carry irrelevant decorations such as horizontal lines, separators, and special characters introduced by various clients. We systematically remove these artefacts. Additionally, we normalise look‐alike characters, such as Unicode homoglyphs, to their standard ASCII counterparts. This normalisation streamlines future text processing and removes elements that could be used to trick detection systems.

The preprocessing of the email body also includes the removal or masking of sensitive information. This includes email addresses, URLs, file paths, and phone numbers, which are replaced with standardised placeholder tags. This approach conceals actual user data while preserving the structural signals (e.g. the presence of a link), essential for phishing detection models. PGP signatures and emojis were also replaced by tags only for preserving structural signals. In addition to the text, we also gather metadata on attachments and hyperlinks. Each attachment's type is recorded, and all links are extracted and analysed for structural complexity.
Finally, recognising that email content may span multiple languages or include code snippets, we segment the cleaned text and apply language detection to each part. These language labels can guide language-specific processing and enrich the feature set used by analytical models. 

After applying the processing pipeline to the combined dataset, some duplicate and empty samples were identified and removed. The resulting dataset, MeAJOR Corpus, comprised a total of 135894 emails.

\section{Baseline Experiments}
\label{sec:experiments}

To analyse the performance of the resulting dataset, we used ML models to assess how the selected features impact classification accuracy. The models RF, XGB, Multilayer Perceptron (MLP), and Convolutional Neural Network (CNN) were selected based on their frequent use and good performance in the reviewed literature. Each model was trained and tested using four different feature combinations to investigate the contribution of different types of information to the overall performance. 

The first feature set, referred to as \textit{Text Only Set}, included only the textual data extracted from the \texttt{subject} and \texttt{body} features of the final dataset. This baseline set provided a foundation to evaluate model performance relying solely on raw text, without incorporating any additional contextual information. Given that phishing emails often exhibit distinctive linguistic patterns, focusing on textual content allowed us to establish an essential reference point for detection effectiveness. 
The second set, referred to as the \textit{Text + URL Set}, combined the \textit{Text Only Set} with URL-based features, namely the total number of URLs, the maximum and average URL length, and the number of subdomains. The third set, called the \textit{Text + Attachment Set}, appended attachment-related features to the \textit{Text Only Set}, namely the total number of attachments, a binary flag indicating their presence, and the top-level MIME types observed. Finally, the \textit{Text + URL + Attachment Set} merge all previous features, offering a holistic representation of each email. This last set was designed to reflect a real-world detection scenario in which multiple phishing indicators may coexist and interact, potentially improving the model's ability to flag deceptive messages.

All the textual features present on all the sets underwent NLP vectorisation. Specifically, we chose the word embedding technique FastText because it incorporates subword (character n-gram) information, allowing rare or unseen words to automatically get embeddings and making morphologically related words closer in vector space. Before applying this technique, the text was lowercased, tokenised, and trimmed of words with little semantic meaning (stopwords and punctuation marks). Once the email content was standardised and tokenised, a numerical vector was generated for each token using a FastText model pre-trained using large amounts of textual data. Since each email contains several tokens, and each token has a corresponding generated vector, we then average all vectors of an email, resulting in a single vector per email content and subject.

Model performance was evaluated using Accuracy, Precision, Recall, and F1-score (F1). To ensure fair and optimised comparisons across all models, hyperparameter tuning was conducted for each one using a 5-fold cross-validation procedure. After tuning, the best hyperparameters were selected and the final models were trained and evaluated using an 80/20 holdout validation split, ensuring consistent evaluation across all feature sets. The performance across all models is summarised in Table \ref{tab:model_results}.

\begin{table}[h!]
    \centering
    \scriptsize
    \caption{Models Results}
    \label{tab:model_results}
    \begin{tabular}{c|c|c|c|c|c}
        \textbf{Model} & \textbf{Feature Set} & \textbf{Acc} & \textbf{Pre} & \textbf{Recall} & \textbf{F1} \\
        \hline
        RF & Text & 96.02 & 97.19 & 94.64 & 95.90 \\
        RF & Text+URL & \textbf{96.73} & \textbf{97.76 }& 95.53 & \textbf{96.63} \\
        RF & Text+Att & 95.86 & 96.99 & 94.52 & 95.74 \\
        RF & Text+URL+Att & 96.70 & 97.70 & \textbf{95.55} & 96.61 \\
        \hline
        XGB & Text & 97.93 & 97.96 & 97.82 & 97.89 \\
        XGB & Text+URL & \textbf{98.37} & \textbf{98.51} & 98.17 & \textbf{98.34} \\
        XGB & Text+Att & 97.85 & 97.87 & 97.75 & 97.81 \\
        XGB & Text+URL+Att & 98.35 & 98.36 & \textbf{98.28} & 98.32 \\
        \hline
        MLP & Text & 97.61 & 97.43 & \textbf{97.72} & 97.57 \\
        MLP & Text+URL & 96.70 & \textbf{97.98} & 95.25 & 96.59 \\
        MLP & Text+Att & \textbf{97.73} & 97.77 & 97.61 & \textbf{97.69} \\
        MLP & Text+URL+Att & 97.22 & 97.32 & 97.01 & 97.17 \\
        \hline
        CNN & Text & 97.49 & \textbf{98.02} & 96.85 & 97.43 \\
        CNN & Text+URL & \textbf{97.69} & 97.42 & \textbf{97.90} & \textbf{97.66} \\
        CNN & Text+Att & 97.59 & 97.28 & 97.83 & 97.55 \\
        CNN & Text+URL+Att & 97.65 & 97.28 & 97.96 & 97.62 \\
    \end{tabular}
    \vspace{1mm}
    \parbox{\linewidth}{\scriptsize
    \textit{Note:} Feature set abbreviations: Text = \texttt{Text Only Set}, Text+URL = \texttt{Text + URL Set}, Text+Att = \texttt{Text + Attachment Set}, Text+URL+Att = \texttt{Text + URL + Attachment Set}}
\end{table}

Analysing the results, it can be noticed that the RF model showed significant improvement when URL features were included alongside textual data, achieving an F1 of 96.63\%. This corroborates findings by Naswir et al.~\cite{Naswir2022} regarding the value of URL features, though our implementation uses a more limited feature set than their hybrid approaches. In contrast, attachment features showed a neutral or slightly negative impact on the overall performance, consistent with Zhang et al.~\cite{Zhang2025} who observed minimal contribution from attachments in their large-scale study. Comparing the \textit{Text + Attachment} and \textit{Text + URL + Attachment} feature sets with their counterparts without the attachment features, confirms that URL information is particularly valuable for phishing detection when using RF, while attachment information has little to no impact.

XGB consistently delivered strong results across all feature combinations, achieving 98.34\% F1 with the \textit{Text + URL Set}. This exceeds Omotehinwa et al.~\cite{Omotehinwa2023} (98.16\% F1) and approaches ensemble methods like Chanis and Arampatzis~\cite{Chanis2024} (98.43\% F1), despite our simpler feature engineering. The robustness of the \textit{Text Only Set} (97.89\% F1) supports Khalid et al.~\cite{Khalid2024} on text-centric efficacy, though our URL augmentation shows greater gains than their stylometric features. Attachment features again demonstrated neutral-to-negative impacts.

The MLP also exhibited strong results across all feature sets, reaching its best performance at 97.69\% F1 with the \textit{Text + Attachment Set}, validating Altwaijry et al.~\cite{Altwaijry2024} on neural networks' text proficiency. Notably, URLs caused a 1.1\% F1 drop, contrasting sharply with ensemble models and highlighting MLP's ineffectiveness with structured features (a limitation not observed in hybrid architectures like the one from Atawneh and Aljehani~\cite{Atawneh2023}).

Finally, the CNN model, evaluated on the same feature sets, achieved its best result when combining textual and URL data, reaching 97.66\% F1 with the \textit{Text + URL Set}. This demonstrated superior structural feature handling versus MLP, though falling short of Kaushik and Rathore's work~\cite{Kaushik2023} (99\% F1), where a specialised URL-to-image transformation is used to capture intricate visual patterns within URLs. The CNN model also maintained high precision and recall across configurations, showing its ability to generalise well even on diverse feature inputs. Notably, adding attachment features showed slightly more positive impacts when compared to the ensemble models, suggesting CNN's adaptability to heterogeneous inputs aligns with Zhang et al.~\cite{Zhang2016} on multi-modal processing.

\section{Discussion}
\label{sec:discussion}

The results obtained from the baseline experiments are crucial for validating the effectiveness, representativeness, and utility of the newly generated dataset presented in this study. Our results robustly confirm the established trend, highlighted by Alhuzali et al.~\cite{Alhuzali2025} and Chanis et al.~\cite{Chanis2024}, that ensemble models such as XGB and RF consistently excel with structured email features. XGB emerged as the top performer in our evaluation, achieving a peak F1 of 98.34\% using the combined \textit{Text + URL Set}. This performance is highly competitive with recent state-of-the-art results, as it surpasses the 98.16\% F1 reported by Omotehinwa et al.~\cite{Omotehinwa2023} using XGB on the Enron-Spam corpus and approaches the 98.43\% F1 achieved by Chanis et al.~\cite{Chanis2024} using a sophisticated stacked ensemble incorporating stylometric features. Importantly, our result was obtained with comparatively simpler feature engineering, underscoring the quality and richness of the underlying dataset. The strong performance of the \textit{Text Only Set} (97.89\% F1) aligns with findings by Khalid et al.~\cite{Khalid2024} on the efficacy of text-centric approaches. However, our integration of URL features presented more substantial gains than the stylometric augmentations reported in their LogiTriBlend model.

The DL models, CNN and MLP, also demonstrated robust performance, particularly with enriched feature sets, supporting findings by Atawneh et al.~\cite{Atawneh2023} and Kaushik et al.~\cite{Kaushik2023} on the applicability of DL to textual features. However, our results reveal the interesting nuance that classical MLP architectures matched or even slightly surpassed CNN performance in some configurations, especially when incorporating structured features like attachments. This contrasts with some state-of-the-art hybrid DL approaches~\cite{Atawneh2023, Kaushik2023}, but highlights a potential advantage of simpler neural architectures when robust feature engineering provides discriminative inputs rather than relying solely on end-to-end representation learning. Our CNN model's peak F1 of 97.66\% with the \textit{Text + URL Set} falls short of Kaushik et al. 99\%~\cite{Kaushik2023}, which leveraged specialised URL-to-image transformations. This gap suggests potential for future work incorporating similar advanced feature representations using our dataset.

The significant performance boost observed when adding URL features aligns strongly with the current state of the art, reinforcing findings by Khalid et al.~\cite{Khalid2024}, Naswir et al.~\cite{Naswir2022}, and Zhang et al.~\cite{Zhang2016} on the critical importance of lexical and structural URL characteristics for identifying phishing attempts. While attachment features showed a more neutral or context-dependent impact in our experiments (consistent with the minimal contribution noted by Zhang et al.~\cite{Zhang2025}), their inclusion reflects the multi-modal learning trend emphasised in cutting-edge research~\cite{Zhang2025, Yoon2024}. Our dataset's design inherently supports this multi-modal approach by providing diverse feature categories.

\section{Conclusion}
\label{sec:conclusion}

This paper presents MeAJOR Corpus, a novel multi-source phishing email dataset designed to overcome critical limitations in existing resources and advance the development of robust ML-based detection systems. By strategically integrating and curating five open-source corpora, we constructed a dataset of 135894 labelled emails. This integration significantly enhances sample diversity, volume, and representativeness, capturing a broader spectrum of phishing tactics and legitimate email structures than isolated or smaller datasets allow.

Our extensive feature engineering constitutes a core strength of the dataset, setting it apart from existing resources. Firstly, by providing both broad, foundational features (like raw body text and URL lists) for downstream custom feature extraction and specific, ready-to-use engineered features (like URL counts and attachment flags), it offers great flexibility. Secondly, it integrates signals across technical, structural, and linguistic dimensions, enabling holistic analysis that surpass the narrow focus of many existing corpora. Thirdly, the consistent preprocessing and anonymisation pipeline ensures feature quality and privacy compliance, addressing inconsistencies found in prior combined datasets.

The MeAJOR Corpus and its feature engineering's utility were validated through experiments with four classification models (RF, XGB, MLP, CNN) across multiple feature configurations. The results were compelling since XGB achieved a peak F1 of 98.34\% using the \textit{Text + URL Set}, closely approaching state-of-the-art benchmarks while utilising significantly simpler feature engineering than comparable studies. This high performance, replicated across models, robustly confirms the dataset's capacity to support accurate and reliable phishing detection. The substantial performance boost observed when incorporating URL features underscores the critical value of the structural URL information captured in our feature set. Furthermore, the strong results obtained with the \textit{Text Only Set} (XGB: 97.89\% F1) affirm the quality of the linguistic data preserved through our preprocessing. While attachment features showed a more context-dependent impact, their inclusion reflects the dataset's inherent support for multi-modal learning, a key trend in advanced detection research.

Future work can leverage the MeAJOR Corpus' diverse features to explore advanced techniques like LLMs, transformer architectures, multi-modal fusion, or synthetic data generation, driving further innovation in safeguarding users against evolving email threats. Exploration of additional data modalities and feature engineering techniques also presents promising avenues.

\section*{Acknowledgment}
This research was funded by the Itea Project VESTA (\#21011) and Project VESTA (NORTE2030-FEDER-00382400) co-funded by Portugal 2030. Furthermore, this work also received funding from the project UIDB/00760/2020.

\bibliographystyle{IEEEtran} 
\bibliography{bibliography}

@misc{IBM_data_breach_report_2024,
    author = {{I}nternational {B}usiness {M}achines Corporation},
    title = {{C}ost of a data breach 2024 | {I}{B}{M}},
    howpublished = {\url{https://www.ibm.com/reports/data-breach}},
    year = {2024},
    note = {accessed: 2025-04-30},
}

@misc{
    apwg_q1,
    author = {{A}nti-{P}hishing {W}orking {G}roup},
    title = {{P}hishing {A}ctivity {T}rends {R}eport 1st {Q}uarter},
    howpublished = {\url{https://docs.apwg.org/reports/apwg_trends_report_q1_2024.pdf}},
    year = {2024},
    note = {accessed: 2025-04-30},
}

@misc{
    apwg_q2,
    author = {{A}nti-{P}hishing {W}orking {G}roup},
    title = {{P}hishing {A}ctivity {T}rends {R}eport 2nd {Q}uarter},
    howpublished = {\url{https://docs.apwg.org/reports/apwg_trends_report_q2_2024.pdf}},
    year = {2024},
    note = {accessed: 2025-04-30},
}

@misc{
    apwg_q3,
    author = {{A}nti-{P}hishing {W}orking {G}roup},
    title = {{P}hishing {A}ctivity {T}rends {R}eport 3rd {Q}uarter},
    howpublished = {\url{https://docs.apwg.org/reports/apwg_trends_report_q3_2024.pdf}},
    year = {2024},
    note = {accessed: 2025-04-30},
}

@misc{
    apwg_q4,
    author = {{A}nti-{P}hishing {W}orking {G}roup},
    title = {{P}hishing {A}ctivity {T}rends {R}eport 4th {Q}uarter},
    howpublished = {\url{https://docs.apwg.org/reports/apwg_trends_report_q4_2024.pdf}},
    year = {2024},
    note = {accessed: 2025-04-30},
}

@article{Alhuzali2025,
    title = {In-Depth Analysis of Phishing Email Detection: Evaluating the Performance of Machine Learning and Deep Learning Models Across Multiple Datasets},
    volume = {15},
    ISSN = {2076-3417},
    url = {https://doi.org/10.3390/app15063396},
    number = {6},
    journal = {Applied Sciences},
    publisher = {MDPI AG},
    author = {A. Alhuzali and A. Alloqmani and M. Aljabri and F. Alharbi},
    year = {2025},
    month = mar,
    pages = {3396}
}

@misc{enron_source,
    author       = {{W. W. Cohen}},
    title        = {{E}nron {E}mail  {D}ataset},
    year         = {2015},
    howpublished = {\url{https://www.cs.cmu.edu/~enron/}},
    note         = {accessed: 2025-05-05}
}

@misc{nazario_source,
    author       = {{J. Nazario}},
    title        = {{N}azario {P}hishing  {C}orpus},
    year         = {2005},
    howpublished = {\url{https://monkey.org/~jose/phishing/}},
    note         = {accessed: 2025-05-05}
}

@misc{spamassassin_source,
    author       = {{Apache SpamAssassin Project}},
    title        = {{S}pam{A}ssassin {P}ublic {M}ail {C}orpus},
    year         = {2006},
    howpublished = {\url{https://spamassassin.apache.org/old/publiccorpus}},
    note         = {accessed: 2025-05-05}
}

@misc{nigerian_source,
  author       = {D. Radev},
  title        = {{CLAIR Collection of Fraud Email}},
  year         = {2008},
  howpublished = {ACL Data and Code Repository, ADCR2008T001},
  note         = {\url{http://aclweb.org/aclwiki}},
}

@misc{trec5source,
  author       = {S. Bayes},
  title        = {{Emails for Spam or Ham Classification (TREC 2005)}},
  howpublished = {\url{https://www.kaggle.com/datasets/bayes2003/emails-for-spam-or-ham-classification-trec-2005}},
  note         = {Accessed: 2025-05-09}
}

@misc{trec6source,
  author       = {{S. Bayes}},
  title        = {{Emails for Spam or Ham Classification (TREC 2006)}},
  howpublished = {\url{https://www.kaggle.com/datasets/bayes2003/emails-for-spam-or-ham-classification-trec-2006}},
  note         = {Accessed: 2025-05-09}
}

@misc{trec7source,
    author       = {{S. Bayes}},
    title        = {{Emails for Spam or Ham Classification (TREC 2007)}},
    howpublished = {\url{https://www.kaggle.com/datasets/bayes2003/emails-for-spam-or-ham-classification-trec-2007}},
    note         = {Accessed: 2025-05-09}
}

@misc{ling_source,
  author       = {I. Androutsopoulos and G. Paliouras and V. Karkaletsis and G. Sakkis and C.D. Spyropoulos and P. Stamatopoulos},
  title        = {{Ling‑Spam} Public Dataset},
  howpublished = {\url{http://www.aueb.gr/users/ion/data/lingspam_public.tar.gz}},
  year         = {2000},
  note         = {Accessed: 2025-05-09}
}

@misc{email4s_source,
    title={Phishing Email Detection},
    url={https://doi.org/10.34740/KAGGLE/DSV/6090437},
    publisher={Kaggle},
    author={S. Chakraborty},
    year={2023}
}

@misc{ceas8source,
    title        = {CEAS 2008 Live Spam Challenge Corpus},
    author       = {G. V. Cormack},
    year         = {2008},
    howpublished = {\url{https://plg.uwaterloo.ca/~gvcormac/ceascorpus/}},
    note         = {Accessed: 2025-05-09}
}

@misc{Lee2025,
    url = {https://doi.org/10.48550/ARXIV.2502.04759},
    author = {C. Lee},
    title = {Enhancing Phishing Email Identification with Large Language Models},
    publisher = {arXiv},
    year = {2025},
    copyright = {Creative Commons Attribution 4.0 International}
}

@inbook{Gascon2018,
    title = {Reading Between the Lines: Content-Agnostic Detection of Spear-Phishing Emails},
    ISBN = {9783030004705},
    ISSN = {1611-3349},
    url = {https://doi.org/10.1007/978-3-030-00470-5_4},
    booktitle = {Research in Attacks,  Intrusions,  and Defenses},
    publisher = {Springer International Publishing},
    author = {H. Gascon and S. Ullrich and B. Stritter and K. Rieck},
    year = {2018},
    pages = {69–91}
}

@article{Gualberto2020,
    author={E. S. Gualberto and R. T. Sousa and T. P. B. Vieira and J. P. C. L. Costa and C. G. Duque},
    journal={IEEE Access},
    title={From Feature Engineering and Topics Models to Enhanced Prediction Rates in Phishing Detection},
    year={2020},
    volume={8},
    number={},
    pages={76368-76385},
    url={https://doi.org/10.1109/ACCESS.2020.2989126}
}

@article{Valecha2022,
    author={R. Valecha and P. Mandaokar and H. R. Rao},
    journal={IEEE Transactions on Dependable and Secure Computing},
    title={Phishing Email Detection Using Persuasion Cues},
    year={2022},
    volume={19},
    number={2},
    pages={747-756},
    url={https://doi.org/10.1109/TDSC.2021.3118931}
}

@article{Gallo2024,
    title = {The human factor in phishing: Collecting and analyzing user behavior when reading emails},
    volume = {139},
    ISSN = {0167-4048},
    url = {https://doi.org/10.1016/j.cose.2023.103671},
    journal = {Computers \&amp; Security},
    publisher = {Elsevier BV},
    author = {L. Gallo and D. Gentile and S. Ruggiero and A. Botta and G. Ventre},
    year = {2024},
    month = apr,
    pages = {103671}
}

@article{Desolda2021,
    title = {Human Factors in Phishing Attacks: A Systematic Literature Review},
    volume = {54},
    ISSN = {1557-7341},
    url = {https://doi.org/10.1145/3469886},
    number = {8},
    journal = {ACM Computing Surveys},
    publisher = {Association for Computing Machinery (ACM)},
    author = {G. Desolda and L. S. Ferro and A. Marrella and T. Catarci and M. F. Costabile},
    year = {2021},
    month = oct,
    pages = {1–35}
}

@article{Mbadiwe2024,
    url = {https://doi.org/10.5281/ZENODO.12651047},
    author = {O. N. Mbadiwe and O. C. Nwokonkwo and A. I. Otuonye and C. O. Ikerionwu and C. Etus},
    title = {Challenges of Data Collection and Preprocessing for Phishing Email Detection},
    publisher = {Novelty Journals},
    year = {2024},
    copyright = {Creative Commons Attribution 4.0 International},
    journal = {{International Journal of Novel Research in Computer Science and Software Engineering}},
    volume = {11},
    pages = {45-59},
    number = {2}
}

@article{AlSubaiey2024,
    title = {Novel interpretable and robust web-based AI platform for phishing email detection},
    volume = {120},
    ISSN = {0045-7906},
    url = {https://doi.org/10.1016/j.compeleceng.2024.109625},
    journal = {Computers and Electrical Engineering},
    publisher = {Elsevier BV},
    author = {A. Al-Subaiey and M. Al-Thani and N. A. Alam and K. F. Antora and A. Khandakar and S. A. U. Zaman},
    year = {2024},
    month = dec,
    pages = {109625}
}

@article{Khalid2024,
    title = {LogiTriBlend: A Novel Hybrid Stacking Approach for Enhanced Phishing Email Detection Using ML Models and Vectorization Approach},
    volume = {12},
    ISSN = {2169-3536},
    url = {https://doi.org/10.1109/access.2024.3518923},
    journal = {IEEE Access},
    publisher = {Institute of Electrical and Electronics Engineers (IEEE)},
    author = {A. Khalid and M. Hanif and A. Hameed and Z. Ashraf and M. M. Alnfiai and S. M. M. Alnefaie},
    year = {2024},
    pages = {193807–193821}
}

@article{Chanis2024,
    title = {Enhancing phishing email detection with stylometric features and classifier stacking},
    volume = {24},
    ISSN = {1615-5270},
    url = {https://doi.org/10.1007/s10207-024-00928-7},
    number = {1},
    journal = {International Journal of Information Security},
    publisher = {Springer Science and Business Media LLC},
    author = {I. Chanis and A. Arampatzis},
    year = {2024},
    month = nov 
}

@article{Atawneh2023,
    title = {Phishing Email Detection Model Using Deep Learning},
    volume = {12},
    ISSN = {2079-9292},
    url = {https://doi.org/10.3390/electronics12204261},
    number = {20},
    journal = {Electronics},
    publisher = {MDPI AG},
    author = {S. Atawneh and H. Aljehani},
    year = {2023},
    month = oct,
    pages = {4261}
}

@article{Kaushik2023,
    title = {Deep Learning Multi-Agent Model for Phishing Cyber-attack Detection},
    volume = {11},
    ISSN = {2321-8169},
    url = {https://doi.org/10.17762/ijritcc.v11i9s.7674},
    number = {9s},
    journal = {International Journal on Recent and Innovation Trends in Computing and Communication},
    publisher = {Auricle Technologies,  Pvt.,  Ltd.},
    author = {P. Kaushik and S. P. S. Rathore},
    year = {2023},
    month = aug,
    pages = {680–686}
}

@article{Zhang2016,
    title = {Two-stage ELM for phishing Web pages detection using hybrid features},
    volume = {20},
    ISSN = {1573-1413},
    url = {https://doi.org/10.1007/s11280-016-0418-9},
    number = {4},
    journal = {World Wide Web},
    publisher = {Springer Science and Business Media LLC},
    author = {W. Zhang and Q. Jiang and L. Chen and C. Li},
    year = {2016},
    month = sep,
    pages = {797–813}
}

@article{Zhang2025,
    title = {A combined feature selection approach for malicious email detection based on a comprehensive email dataset},
    volume = {8},
    ISSN = {2523-3246},
    url = {https://doi.org/10.1186/s42400-024-00309-6},
    number = {1},
    journal = {Cybersecurity},
    publisher = {Springer Science and Business Media LLC},
    author = {H. Zhang and Y. Shi and M. Liu and L. Chen and S. Wu and Z. Xue},
    year = {2025},
    month = feb 
}

@misc{Koide2024,
    url = {https://doi.org/10.48550/ARXIV.2402.18093},
    author = {T. Koide and N. Fukushi and H. Nakano and D. Chiba},
    title = {ChatSpamDetector: Leveraging Large Language Models for Effective Phishing Email Detection},
    publisher = {arXiv},
    year = {2024},
    copyright = {Creative Commons Attribution 4.0 International}
}

@article{Tawil2024,
    title = {Comparative Analysis of Machine Learning Algorithms for Email Phishing Detection Using TF-IDF,  Word2Vec,  and BERT},
    volume = {81},
    ISSN = {1546-2226},
    url = {https://doi.org/10.32604/cmc.2024.057279},
    number = {2},
    journal = {Computers,  Materials \&amp; Continua},
    publisher = {Tech Science Press},
    author = {A. A. Tawil and L. Almazaydeh and D. Qawasmeh and B. Qawasmeh and M. Alshinwan and K. Elleithy},
    year = {2024},
    pages = {3395–3412}
}

@misc{Yasin2016,
  url = {https://doi.org/10.48550/ARXIV.1608.02196},
  author = {A. Yasin and A. Abuhasan},
  title = {An intelligent classification model for phishing email detection},
  publisher = {arXiv},
  year = {2016},
  copyright = {arXiv.org perpetual,  non-exclusive license}
}

@article{Mieronkoski2020,
    title = {Developing a pain intensity prediction model using facial expression: A feasibility study with electromyography},
    volume = {15},
    ISSN = {1932-6203},
    url = {https://doi.org/10.1371/journal.pone.0235545},
    number = {7},
    journal = {PLOS ONE},
    publisher = {Public Library of Science (PLoS)},
    author = {R. Mieronkoski and E. Syrj\"{a}l\"{a} and M. Jiang and A. Rahmani and T. Pahikkala and P. Liljeberg and S. Salanter\"{a}},
    editor = {A. Jones},
    year = {2020},
    month = jul,
    pages = {e0235545}
}

@misc{Joshi2019,
    url = {https://doi.org/10.48550/ARXIV.1910.06277},
    author = {A. Joshi and L. Lloyd and P. Westin and S. Seethapathy},
    title = {Using Lexical Features for Malicious URL Detection -- A Machine Learning Approach},
    publisher = {arXiv},
    year = {2019},
    copyright = {arXiv.org perpetual,  non-exclusive license}
}

@article{Ahmed2022,
    title = {Malicious URL Detection Using Decision Tree-based Lexical Features Selection and Multilayer Perceptron Model},
    volume = {6},
    ISSN = {2521-4209},
    url = {https://doi.org/10.21928/uhdjst.v6n2y2022.pp105-116},
    number = {2},
    journal = {UHD Journal of Science and Technology},
    publisher = {University of Human Development},
    author = {W. F. Ahmed and N. G. M. Jameel},
    year = {2022},
    month = nov,
    pages = {105–116}
}

@article{Xie2022,
    title = {Phishing short URL detection based on link jumping on social networks},
    volume = {47},
    ISSN = {2271-2097},
    url = {https://doi.org/10.1051/itmconf/20224701009},
    journal = {ITM Web of Conferences},
    publisher = {EDP Sciences},
    author = {B. Xie and Q. Li and N. Wei},
    editor = {L. Nguyen},
    year = {2022},
    pages = {01009}
}

@article{Chen2023,
    title = {Malicious Office Macro Detection: Combined Features with Obfuscation and Suspicious Keywords},
    volume = {13},
    ISSN = {2076-3417},
    url = {https://doi.org/10.3390/app132212101},
    number = {22},
    journal = {Applied Sciences},
    publisher = {MDPI AG},
    author = {X. Chen and  W. Wang and W. Han},
    year = {2023},
    month = nov,
    pages = {12101}
}

@misc{Rudd2018,
  url = {https://doi.org/10.48550/ARXIV.1804.08162},
  author = {E. M. Rudd and R. Harang and J. Saxe},
  title = {MEADE: Towards a Malicious Email Attachment Detection Engine},
  publisher = {arXiv},
  year = {2018},
  copyright = {arXiv.org perpetual,  non-exclusive license}
}

@article{Kulkarni2020,
    title = {Effect of Header-based Features on Accuracy of Classifiers for Spam Email Classification},
    volume = {11},
    ISSN = {2158-107X},
    url = {https://doi.org/10.14569/ijacsa.2020.0110350},
    number = {3},
    journal = {International Journal of Advanced Computer Science and Applications},
    publisher = {The Science and Information Organization},
    author = {P. Kulkarni and J. R. Saini and H. Acharya},
    year = {2020}
}

@article{Yoon2024,
    title = {Phishing Webpage Detection via Multi-Modal Integration of HTML DOM Graphs and URL Features Based on Graph Convolutional and Transformer Networks},
    volume = {13},
    ISSN = {2079-9292},
    url = {https://doi.org/10.3390/electronics13163344},
    number = {16},
    journal = {Electronics},
    publisher = {MDPI AG},
    author = {J. Yoon and S. Buu and H. Kim},
    year = {2024},
    month = aug,
    pages = {3344}
}

@article{Yoon2025,
    title = {Reinforced Disentangled HTML Representation Learning with Hard-Sample Mining for Phishing Webpage Detection},
    volume = {14},
    ISSN = {2079-9292},
    url = {https://doi.org/10.3390/electronics14061080},
    number = {6},
    journal = {Electronics},
    publisher = {MDPI AG},
    author = {J. Yoon and S. Buu and H. Kim},
    year = {2025},
    month = mar,
    pages = {1080}
}

@inproceedings{Hao2016,
    series = {CCS’16},
    title = {PREDATOR: Proactive Recognition and Elimination of Domain Abuse at Time-Of-Registration},
    url = {https://doi.org/10.1145/2976749.2978317},
    booktitle = {Proceedings of the 2016 ACM SIGSAC Conference on Computer and Communications Security},
    publisher = {ACM},
    author = {S. Hao and A. Kantchelian and B. Miller and V. Paxson and N. Feamster},
    year = {2016},
    month = oct,
    pages = {1568–1579},
    collection = {CCS’16}
}

@article{Chiba2025,
    title = {DomainDynamics: Advancing lifecycle-based risk assessment of domain names},
    volume = {153},
    ISSN = {0167-4048},
    url = {https://doi.org/10.1016/j.cose.2025.104366},
    journal = {Computers \&amp; Security},
    publisher = {Elsevier BV},
    author = {D. Chiba and H. Nakano and T. Koide},
    year = {2025},
    month = jun,
    pages = {104366}
}

@unpublished{afonso2025rethinking,
    title        = {Rethinking Phishing Detection: How Dataset Quality Affects Model Generalization},
    author       = {P. Afonso and E. Maia and I. Amorim and I. Praça},
    year         = {2025},
    note         = {Manuscript in preparation},
}

@article{Naswir2022,
    title = {THE EFFECTIVENESS OF URL FEATURES ON PHISHING EMAILS CLASSIFICATION USING MACHINE LEARNING APPROACH},
    volume = {11},
    ISSN = {2289-2192},
    url = {https://doi.org/10.17576/apjitm-2022-1102-04},
    number = {02},
    journal = {Asia-Pacific Journal of Information Technology and Multimedia},
    publisher = {Penerbit Universiti Kebangsaan Malaysia (UKM Press)},
    author = {A. F. Naswir and L. Q. Zakaria},
    year = {2022},
    month = dec,
    pages = {49–58}
}

@article{Omotehinwa2023,
  title = {Hyperparameter Optimization of Ensemble Models for Spam Email Detection},
  volume = {13},
  ISSN = {2076-3417},
  url = {https://doi.org/10.3390/app13031971},
  number = {3},
  journal = {Applied Sciences},
  publisher = {MDPI AG},
  author = {T. O. Omotehinwa and D. O. Oyewola},
  year = {2023},
  month = feb,
  pages = {1971}
}

@article{Altwaijry2024,
    title = {Advancing Phishing Email Detection: A Comparative Study of Deep Learning Models},
    volume = {24},
    ISSN = {1424-8220},
    url = {https://doi.org/10.3390/s24072077},
    number = {7},
    journal = {Sensors},
    publisher = {MDPI AG},
    author = {N. Altwaijry and I. Al-Turaiki and R. Alotaibi and F. Alakeel},
    year = {2024},
    month = mar,
    pages = {2077}
}

@article{Salloum2022,
    title = {A Systematic Literature Review on Phishing Email Detection Using Natural Language Processing Techniques},
    volume = {10},
    ISSN = {2169-3536},
    url = {https://doi.org/10.1109/ACCESS.2022.3183083},
    journal = {IEEE Access},
    publisher = {Institute of Electrical and Electronics Engineers (IEEE)},
    author = {Salloum,  Said and Gaber,  Tarek and Vadera,  Sunil and Shaalan,  Khaled},
    year = {2022},
    pages = {65703-65727}
}

@article{Doshi2023,
    title = {A comprehensive dual-layer architecture for phishing and spam email detection},
    volume = {133},
    ISSN = {0167-4048},
    url = {https://doi.org/10.1016/j.cose.2023.103378},
    journal = {Computers \&amp; Security},
    publisher = {Elsevier BV},
    author = {Doshi,  Jay and Parmar,  Kunal and Sanghavi,  Raj and Shekokar,  Narendra},
    year = {2023},
    month = oct,
    pages = {103378}
}

@article{WooyunXSSDataset,
    title={Providing Email Privacy by Preventing Webmail from Loading Malicious XSS Payloads},
    author={Fang, Yong and Xu, Yijia and Jia, Peng and Huang, Cheng},
    journal={Applied Sciences},
    volume={10},
    number={13},
    pages={4425},
    year={2020},
    publisher={Multidisciplinary Digital Publishing Institute}
}

@misc{Xiang2025,
    url = {https://doi.org/10.48550/arxiv.2505.09999},
    author = {Y. Xiang and X. Li and K. Qian and W. Yu and E. Zhai and X. Jin},
    title = {ServeGen: Workload Characterization and Generation of Large Language Model Serving in Production},
    publisher = {arXiv},
    year = {2025},
    copyright = {Creative Commons Attribution Share Alike 4.0 International}
}

@article{vinayakumar2018deepanti,
    title={DeepAnti-PhishNet: applying deep neural networks for phishing email detection},
    author={R. Vinayakumar and H. B. B. Ganesh and M. A. Kumar and K. P. Soman},
    journal={CEN-AISecurity@ IWSPA},
    pages={40-50},
    year={2018}
}

@article{Fang2019,
    title={Phishing email detection using improved RCNN model with multilevel vectors and attention mechanism},
    author={Y. Fang and C. Zhang and C. Huang and L. Liu and Y. Yang},
    journal={IEEE Access},
    volume={7},
    pages={56329-56340},
    year={2019},
    publisher={IEEE},
    url = {https://doi.org/10.1109/ACCESS.2019.2913705}
}

@misc{PhishTank,
    title = {{PhishTank: The Phishing Archive}},
    howpublished = {\url{https://www.phishtank.org/}},
    note = {accessed: 2025-06-26},
    organization = {OpenDNS}
}

@misc{AlexaTopSites,
    title = {{Alexa Top Sites}},
    howpublished = {\url{https://web.archive.org/web/*/https://www.alexa.com/topsites}},
    note = {Accessed via Wayback Machine, as Alexa.com's public Top Sites list service was discontinued. Original service by Alexa Internet, an Amazon company.},
    year = {1996-2022}
}

@inproceedings{Kang2007,
    title={Advanced white list approach for preventing access to phishing sites},
    author={J. Kang and D. Lee},
    booktitle={2007 International Conference on Convergence Information Technology (ICCIT 2007)},
    pages={491-496},
    year={2007},
    organization={IEEE},
    url = {https://doi.org/10.1109/ICCIT.2007.50}
}

@inproceedings{Prakash2010,
    title = {Phishnet: predictive blacklisting to detect phishing attacks},
    author = {P. Prakash and M. Kumar and R. R. Kompella and M. Gupta},
    booktitle = {2010 Proceedings IEEE INFOCOM},
    pages = {1-5},
    year = {2010},
    organization = {IEEE},
    url = {https://doi.org/10.1109/INFCOM.2010.5462216}
}

@article{Juneja2014,
    title={A Survey on Email Spam Types and Spam Filtering Techniques},
    author={Juneja, Prachi Goyal and Pateriya, RK},
    journal={International Journal of Engineering Research},
    volume={3},
    number={3},
    year={2014}
}

@article{Kyaw2024,
    title = {A Systematic Review of Deep Learning Techniques for Phishing Email Detection},
    volume = {13},
    ISSN = {2079-9292},
    url = {https://doi.org/10.3390/electronics13193823},
    number = {19},
    journal = {Electronics},
    publisher = {MDPI AG},
    author = {P. H. Kyaw and J. Gutierrez and A. Ghobakhlou},
    year = {2024},
    month = sep,
    pages = {3823}
}

\end{document}